\documentclass[a4paper]{article}

\usepackage{INTERSPEECH2021}
\usepackage{booktabs}
\usepackage{multirow}
\usepackage{blindtext}
\pdfoutput=1
\title{SRIB-LEAP   submission to Far-field Multi-Channel \\  Speech Enhancement Challenge for Video Conferencing}
\name{R G Prithvi Raj$^1$, Rohit Kumar$^2$, M K Jayesh$^1$, Anurenjan Purushothaman$^2$, Sriram Ganapathy$^2$, and M A Basha Shaik$^1$ }
\address{
  $^1$Samsung Research \& Development Institute India, Bangalore\\
  $^2$LEAP lab, Electrical Engineering, Indian Institute of Science, Bangalore, India.}
\email{\{rohitk,anurenjanr,sriramg\}@iisc.ac.in, \{p.gudepu,jayesh.mk,m.shaik\}@samsung.com}

\begin{document}

\maketitle
\begin{abstract}
This paper presents the details of the SRIB-LEAP submission to the ConferencingSpeech challenge $2021$. The challenge involved the task of multi-channel speech enhancement to improve the quality of far field speech from microphone arrays in a video conferencing room. We propose a two stage method involving a beamformer followed by single channel enhancement. For the beamformer, we incorporated self-attention mechanism as inter-channel processing layer in the filter-and-sum network (FaSNet), an end-to-end time-domain beamforming system. The single channel speech enhancement is done in log spectral domain using convolution neural network  (CNN)-long short term memory (LSTM) based architecture. We achieved improvements in objective quality metrics - perceptual evaluation of speech quality (PESQ) of $0.5$ on the noisy data. On subjective quality evaluation, the proposed approach improved the mean opinion score (MOS) by an absolute measure of $0.9$ over the noisy audio. 
\end{abstract}
\noindent\textbf{Index Terms}: Speech Enhancement, PESQ, STOI, Si-SNR.

\section{Introduction}

Video conferencing plays a very crucial part in day-to-day social interactions. However, video conferencing with far-field microphones in the presence of other noise sources degrades the audio quality and impedes the efficiency of the voice communication. In order to facilitate the development of algorithms for quality improvement in video conferencing setting, ConferencingSpeech $2021$ challenge \cite{conferencingspeech} was initiated. This challenge provides a platform for bench-marking speech enhancement techniques recorded from real speakers recorded in meeting room settings of varying sizes.  

The conventional method of processing the multi-channel audio signal involves the spatial filtering performed via beamforming ~\cite{van1988beamforming,krim1996two}. The method of beamforming performs a delayed and weighted summation of the multiple spatially separated microphones to provide an enhanced audio signal. The advancements to the basic beamforming using blind reference-channel selection and two-step time delay of arrival (TDOA) estimation with Viterbi post processing has been proposed to improve the beamforming algorithm ~\cite{anguera2007acoustic}. An alternate approach to beamforming using a generalized eigen value (GEV) formulation~\cite{warsitz2007blind} involves a spatial filtering in the complex short-time Fourier transform (STFT) domain. The filter is derived by solving an eigen value problem that maximizes the variance in the ``signal'' direction while minimizing the variance in the ``noise'' direction~\cite{warsitz2007blind} or by keeping the variance in the target direction to be unity while minimizing the variance in the other directions (minimum variance distortionless response (MVDR) beamforming)~\cite{higuchi2016robust}. Recently, unsupervised DNN mask estimator based beamforming was also proposed for GEV based beamforming \cite{rohit}.

Even after beamforming of multi channel speech,  the resultant speech signal contains noise and reverberation artifacts, which can further be reduced. The speech enhancement based on neural networks has made noticeable progress in the recent years. The early works by Xu et. al. \cite{xu2014regression} targeted the enhancement of signals corrupted by additive noise where a supervised neural network method was proposed to enhance speech by means of finding a mapping function the noisy signal to the clean speech. In a similar manner, speech separation (the problem of separating the target speaker speech from the background interference) has seen considerable progress using neural methods with ideal ratio mask based mapping \cite{wang2018supervised}.
For reverberant speech, Zhao et al., proposed a LSTM model to predict late reflections in the spectrogram domain \cite{zhao2018late}. A spectral mapping approach using the log-magnitude inputs was attempted by Han et. al \cite{han2014learning}. A mask based approach to dereverberation on the complex short-term Fourier transform domain was explored by Williamson et. al \cite{williamson2017time}. A recurrent neural network model to predict the spectral magnitudes for dereverberation of speech was also proposed by Santos et. al \cite{santos2018speech}. 
Speech enhancement for speech recognition based on neural networks has been explored in \cite{wollmer2013feature,chen2015speech,ganapathy20183,purushothaman2020deep}. 

In our proposed work for the ConfercingSpeech challenge, we have used a combination of FaSNet beamformer model ~\cite{luo2019FaSNet} and a single channel enhancement model using convolutional long short term memory (CLSTM)~\cite{Purushothaman2020}. The filter-and-sum network (FaSNet)~\cite{FaSNetplustcn}, a time-domain filter based beamforming approach, first learns frame-level time-domain beamforming filters.  The model then calculates the filters for all remaining channels  and the filtered outputs at all channels are summed to generate the final output. The novel contribution of our work is the incorporation of the self-attention mechanism (A-FaSNet) to process data along the channel dimension in place of the transform-average-concatenate (TAC) module in \cite{FaSNetplustcn}.

Following the beamforming, the single channel speech signal is  dereverberated using CNN-LSTM model. The single channel enhancement is a variant of work done by Purushothaman et. al.  \cite{Purushothaman2020}. The rest of the paper is organized as follow. Sec.~\ref{pro} describes the systems that is being used in the challenge. The experiments and results on ASR tasks are reported in  Sec.~\ref{exp}. A summary of the work is given in Sec.~\ref{sec:summary}. 

\begin{figure*}[t!]
    \centering
     \includegraphics[scale=0.65]{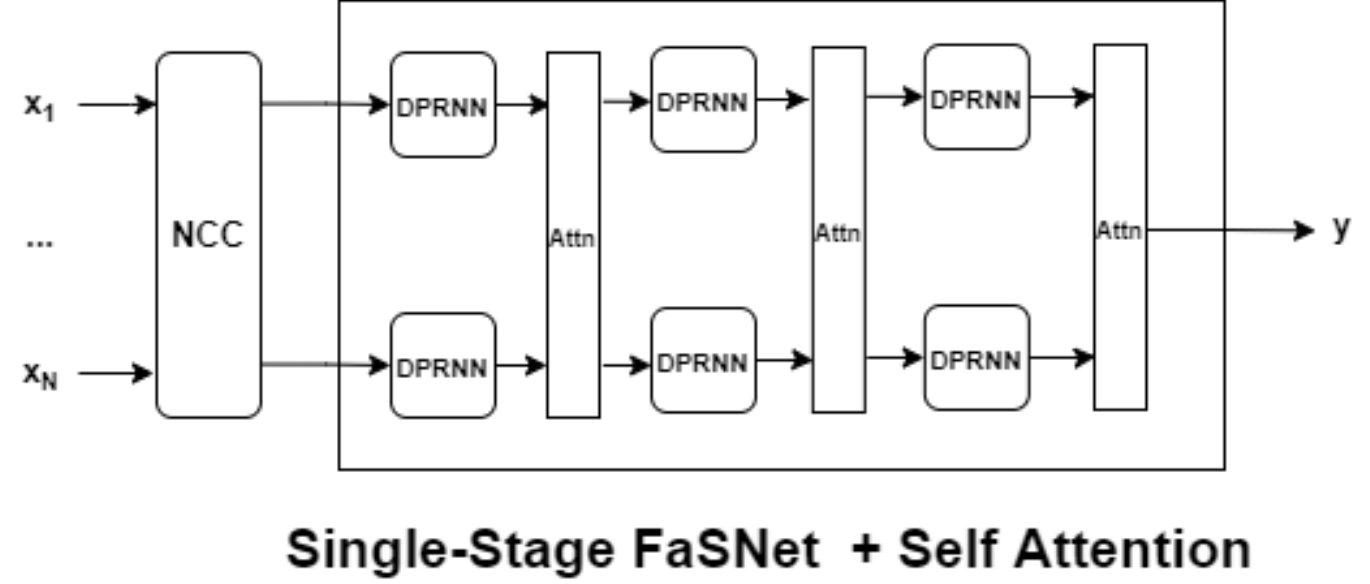}
    \includegraphics[scale=0.65]{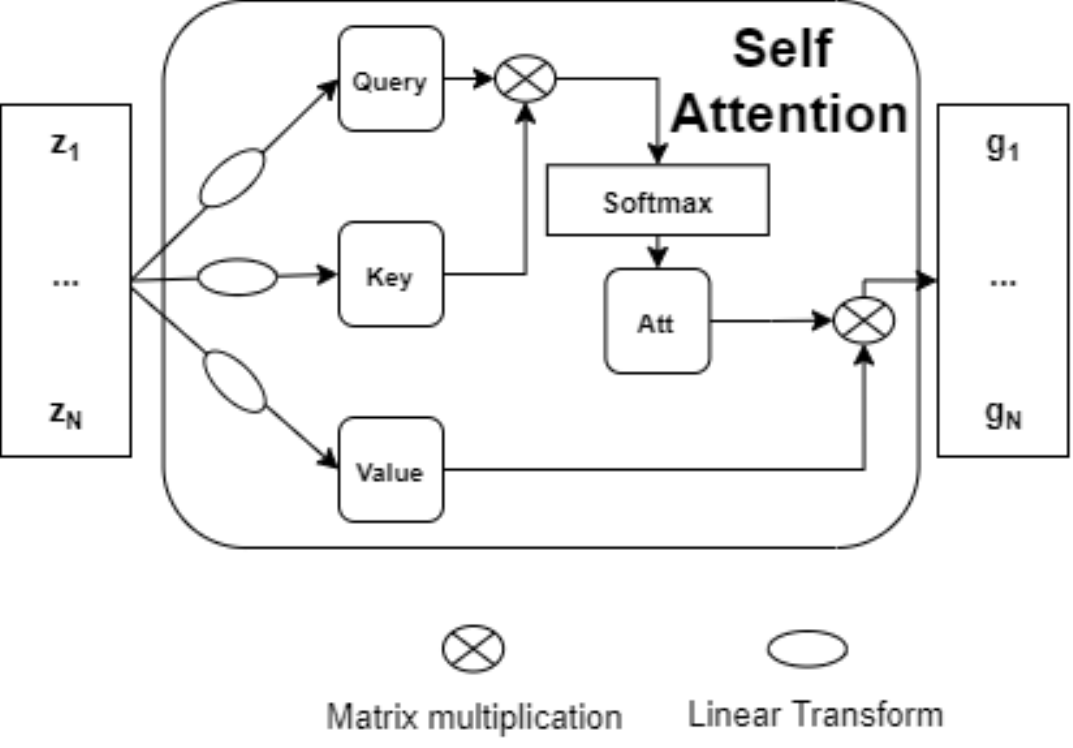}
    \caption{\small (left)  Single stage FaSNet with self-attention (b) Self-attention module. Here, NCC corresponds to normalized cross correlation and DPRNN corresponds to dual path recurrent neural network.   }
    \label{fig:FaSNet}
\end{figure*} 

\section{Proposed system}\label{pro}
The proposed approach consists of two stages - (a) Beamforming based on attention FaSNet model (A-FaSNet) model, (b) Single channel neural enhancement. \footnote{Contact authors: Rohit Kumar, Jayesh MK.}

\subsection{A-FaSNet Model}

FaSNet is a time-domain filter-and-sum
neural beamformer that uses trained filters to denoise and combine multichannel audio.
In this work, inspired by the single-stage FaSNet with TAC architecture proposed in \cite{FaSNetplustcn}, we propose using a self-attention layer in the FaSNet model (shown in Fig.~\ref{fig:FaSNet}).  

The proposed architecture has three main components, (i) context compression, (ii) DPRNN (Dual Path Recurrent Neural Network), and (iii) context decompression.

\subsubsection{Context Compression}
The input multi-channel speech passes through an encoder network which consists of a $1$-$D$ convolution network followed by a RNN layer.
We considered audio signals of length $4$ sec for training and testing, and the sampling frequency used is $16$kHz. The number of channels is fixed as $8$.  The input data of dimension $8\times 64000$ (\# Channels $\times$ \# samples per channel) is passed to a $1$-$D$ convolution block which has $64$ convolution kernels of size $256$.  The stride used is $128$. This encoder network converts the data into a tensor of dimension $(8\times 64\times 502)$. For each frame of the encoded data, four contextual frames from either side are added, making the encoder output dimension $(8\times 64\times 5\times 502)$. By fixing the first encoded channel as reference,  normalized cross correlation (NCC) is calculated between the reference channel and the other channels.  The encoded multi-channel data is passed through a LSTM network and appended with NCC features. The resulting embeddings are passed to the next phase of the model.

\subsubsection{DPRNN With Self-Attention Layer}
The embeddings from different channels are passed through a DPRNN layer (as shown in Fig. \ref{fig:FaSNet}\,(left)).
The DPRNN splits a sequential input into chunks with overlaps.

The A-FaSNet model contains attention layers between all the DPRNN layers  for inter-channel processing.  
We tailor the attention scheme similar to the one proposed in  \cite{attentionbeam2020} which used multi thread dot product attention. The attention layer combines DPRNN states of different channels. 

 Let $Z=[z_0,...,z_{T-1}]$ denote the input tensor, where
$z_t \in R^{h_{size}}$
 is the DPRNN state for time $t$ while $C$ denotes  the number of channels and hidden layer size. Let 
the number of attention heads be $D$. For each attention head,
the input features are transformed into query ($Q$), key ($K$), and
value ($V$) embedding sub-spaces of dimension E as follows:
\begin{eqnarray*}
    Q^{i}_t = W^i_Qz_t + b^i_Q \\
     K^{i}_t = W^i_Kz_t + b^i_K\\
     V^{i}_t = W^i_Vz_t + b^i_V
\end{eqnarray*}
The matrices $Q^i_t$, $K^i_t$ and, $V^i_t$ denote the
query, key, and value matrices, respectively, for the $i^{th}$ attention head at time $t$.  Within each attention head, a product of the query and key matrices is computed which gives the cross-channel similarity. At the output of each attention head, a soft-max layer is applied.  The attention matrix is given as,
\[   A^{i}_t = \text{softmax}((Q^i_t)^T\,K^i_t) \]
The value matrix  is multiplied by  to give attention matrix give the inter mediate output $y_i$,
\[y^i_t=V^{i}_t(A^{i}_t)^T\]
Each $y_i$  is passed through a fully connected network with rectified linear unit (ReLU) activation to generate the attention layer's output $z_i$.

\begin{figure*}[t!]
  \centering
  \includegraphics[scale=0.30]{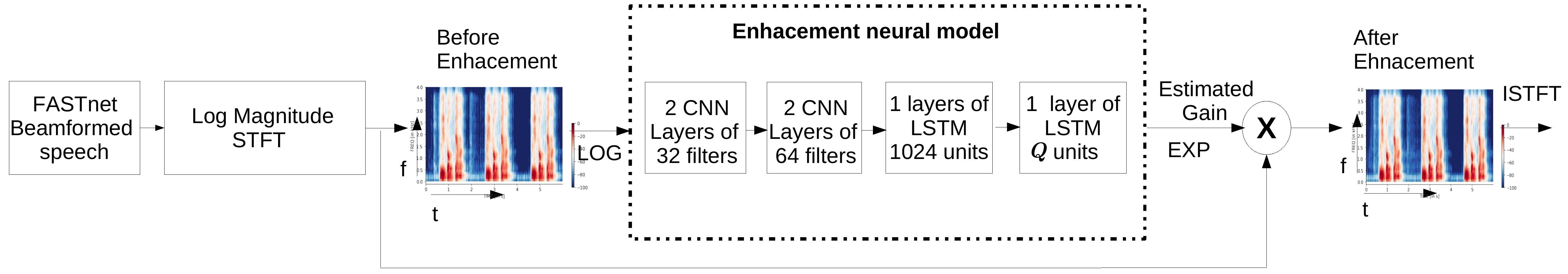}
      \caption{Block schematic of the  single channel neural enhancement model. }
  \label{fig:fig2}
\end{figure*}

\subsubsection{Context Decompression}

The outputs of each DPRNN channel are overlapped and added and passed through the LSTM layer, followed by applying a $1$-D convolution. These operations will result in tensor which acts as a beamforming filter as it is multiplied with the context encoded data of the reference channel to give a beamformed single channel data. This tensor is passed through a decoder block consisting of  $1$-D transposed convolution \cite{1d_trans_conv}
which will convert the tensor $[1, 64, 502]$ to $[1,64000]$ array. This is the final beamformed signal.

\subsubsection{Training of The Model}

The training target is kept as non-reverberant clean speech. The model is trained using the Scale Invariant Signal to Noise Ratio (SI-SNR) cost function~\cite{FaSNetplustcn}.





\subsection{Single Channel Neural  Speech Enhacement}

Let $z(t)$ denote the output of the A-FaSNet beamformed signal. The clean reference signal is denoted by $x(t)$. Let $\textbf{X}(k,n)$ and $\textbf{Z}(k,n)$ denote the short time Fourier transform (STFT) of the clean reference and the beamformed signal respectively, where $k$ denotes the frequency bin index and $n$ denotes the frame index. Then,
\begin{equation}\label{enh_eq}
    \textbf{R}(k,n) = log|{\textbf{X}(k,n)}| - log|{\textbf{Z}(k,n)}|,
\end{equation}
where $\textbf{R}(k,n)$ is the log magnitude residual STFT. 

In the enhancement model shown in Fig.~\ref{fig:fig2}, the input is the log magnitude STFT of beamformed signal. Once this prediction is achieved, the estimate $\textbf{\^{R}}(k,n)$ can be added  to the beamformed signal $\textbf{Z}(k,n)$ to obtain the clean reference signal i.e., $\textbf{\^{X}}(k,n)$. A similar analogy   is the spectral subtraction model where the noise and clean power spectral density (PSD) gets added in noisy speech PSD. If Gaussian assumptions are made for PSD components \cite{martin2005speech}, the Wiener filtering approach to noisy speech enhancement provides the minimum mean squared error, where the noisy PSD is multiplied by the gain of the filter. In a similar manner,  we pose the single channel enhancement problem as an gain estimation problem. The gain in this case is the ratio of the magnitude STFT of clean signal to the magnitude STFT of beamformed signal. This gain estimation is achieved using a deep neural network model having a CLSTM architecture. Following the model training, the speech enhancement is achieved by multiplying (adding) the estimated gain with the (log) magnitude STFT of the beamformed signal. 

\section{Experiments and Results}\label{exp}

\subsection{Dataset}

The clean training speech data is derived from four open source
speech databases: AISHELL-1 \cite{aishell-1}, AISHELL-3 \cite{aishell-3}, VCTK \cite{vctk}, and  Librispeech  \cite{librispeech}. The speech utterances with SNR larger than $15$ dB are selected for training. The total duration of clean training speech is around $550$ hours. The image method is performed to simulate room impulse response (RIR). The room size ranges from $3 \times 3 \times 3$ m$^3$ to $8 \times 8 \times 8$ m$^3$. The microphone array is randomly placed in the room with height ranges from $1.0$ to $1.5$ $m$. This generates a total of around $20,000$ RIRs.

The development test set could be categorized into three parts: simulated recordings, semi-real recordings, and real speaker recordings.  The $1,624$ clean speech recordings  selected from AISHELL-1, AISHELL-3, and VCTK and the $800$ noise samples selected from music and speech noise (MUSAN) corpus are
used for the generation of simulation recordings. The semi-real recordings
consists of $2.35$ hours of playback English speech segments and
$2.31$ hours of real speaker’s Chinese speech segments. There are more than $200$ real recordings, which are from
$12$ real speakers and their ages range from $18-50$ years old.

\begin{table*}[h!]

\centering
    \caption{Comparison of various objective speech evaluation metrics across models.\label{tab:pesq}}
\begin{tabular}{@{}cccccc@{}}
\toprule
Dev set (simulated data) & Array &  & PESQ & STOI & Si SNR \\ \midrule
\multirow{9}{*}{Single MA} & \multirow{4}{*}{Circular} & Noisy & 1.514 & 0.824 & 4.566 \\ \cmidrule(l){3-6} 
 &  & Enhanced (baseline) & 1.990 & 0.888 & 9.248 \\ \cmidrule(l){3-6} 
 &  &  A-FaSNet  (Proposed method) &1.841 &0.853 &7.71   \\ \cmidrule(l){3-6} 
 &  &   A-FaSNet + Neural Enhancement (Proposed method) &1.972  &0.863  &7.71  \\ \cmidrule(l){2-6} 
 & \multirow{3}{*}{Linear uniform} & Noisy & 1.534 & 0.829 & 4.720 \\ \cmidrule(l){3-6} 
 &  & Enhanced (baseline) & 2.035 & 0.893 & 4.720 \\ \cmidrule(l){3-6} 
 &  &   A-FaSNet  (Proposed method) &1.863  &0.857  &7.89  \\ \cmidrule(l){3-6}
 &  &  A-FaSNet + Neural Enhancement (Proposed method) &1.995  &0.867  & 7.88   \\ \cmidrule(l){2-6} 
 & \multirow{3}{*}{Linear nonuniform} & Noisy & 1.515 & 0.823 & 4.475 \\ \cmidrule(l){3-6} 
 &  & Enhanced (baseline) & 1.999 & 0.888 & 9.159 \\ \cmidrule(l){3-6}
 &  &   A-FaSNet  (Proposed method) &1.850  &0.854  &7.74   \\ \cmidrule(l){3-6} 
 &  &  A-FaSNet + Neural Enhancement (Proposed method)  &1.984  &0.864   &7.74   \\ \bottomrule
\end{tabular}
\end{table*}

\begin{table}[!htbp]
\centering
\caption{The subjective evaluation results on Eval Data set for the proposed method (submitted for the challenge), the baseline system and the noisy signal in terms of mean opinion score (MOS), S-MOS, dMOS, N-MOS and, CI .\label{tab:mos}} 


\begin{tabular}{@{}c|cccc@{}}
\toprule
Eval set (real data) & MOS  & S-MOS & N-MOS & $95\%$CI   \\ \hline
Noisy & 2.56 & 2.93  & 3.03  &  0.02 \\
Baseline & 3.43 & 3.55  & 3.48  &  0.03 \\
Proposed & \textbf{3.45} & \textbf{3.38}  & \textbf{3.36}  & \textbf{0.04}   \\
\bottomrule
\end{tabular}
\vspace{-3mm}
\end{table}

\subsection{Training of A-FaSNet model}
The single-stage fasnet model is trained as described in \cite{FaSNetplustcn} with self-attention replacing the TAC block.
Normalized cross-correlation (NCC) between channels is used as inter-channel feature and is fed to the model. The window size of $16$ms with $16$ms of context on either side is used for each frame with frame shift of 8ms.

Though the FaSNet model is used for multi speaker speech separation, for the enhancement task we use single speaker output.  
Self attention layer implemented across channel dimension has 64 dimensional embedding spaces and 2 attention heads.  

 The model is trained for 30 epochs with learning rate set to 0.0001 and weight decay to 1e-5. We use Adam optimization and a batch size of 4 waveforms with zero padding of the inputs, implemented in PyTorch version 1.2.0.


\subsection{Training of Enhancement Network}

The block schematic of the enhancement model is shown in Fig.~\ref{fig:fig2}.  The input to the enhancement model is the log magnitude STFT of beamformed speech. The model is trained to learn the  residual gain which is the ratio of the clean magnitude STFT (clean speech signal) with the beamformed signal magnitude STFT. For training, for every noisy audio file, we have access to its corresponding clean audio file in simulated setting. At the input, we use the beamformed output from the FaSNet architecture and perform STFT with Hanning window  of size $320$ samples  with an overlap of $160$ samples, and perform $512$ point STFT on those samples. 

The architecture of the neural model is based on convolutional long short term memory (CLSTM) networks   with $3.2$ million parameters. The input $2$-D data of sub-band envelopes are fed to a set of convolutional layers where the first two layers have $32$ filters each with kernels of size of $5 \times 21$. The next two CNN layers have $64$ filters with $3 \times 41$ kernel size. All the CNN layer outputs with ReLU activation are zero padded to preserve the input size and no pooling operation is performed.  The output of the CNN layers are reshaped to perform time domain recurrence using $2$ layers of LSTM cells. The first LSTM layers have $512$ cells while the last layer has $257$ cells corresponding to the size of the target signal. The training criteria is based on the mean square error between the target and predicted output. The model is trained with stochastic gradient descent using Adam optimizer. These enhanced magnitude STFT is combined with the original phase of the beamformed signal to perform inverse STFT. This generates the time domain audio signal back.

\subsection{Performance Metrics}
The main aspects of interest for speech enhancement is
quality and intelligibility. The speech quality is largely subjective \cite{deller2000discrete} and can be defined as the result of the judgement based on the characteristics that allow to perceive speech according to the expectations of a listener. The intelligibility can be considered a more objective attribute, because it refers to the speech content \cite{loizou2013speech}.
We use MOS (Mean Opinion Score) for subjective scoring which is a $5$ point scale ranging $1$ (highly distorted) to $5$(excellent). 

The MOS results are tabulated in  Table\,\ref{tab:mos}. The table contains the following subjective measures:
\begin{itemize}
\item MOS: Determination of subjective global MOS. 
\item S-MOS: Determination of subjective speech MOS (S-MOS) where the listener is asked to attend only to the speech signals.
\item N-MOS: Determination of subjective noise MOS (N-MOS) where the listener is asked to attend only to the background.
 
\end{itemize} 
Table~\ref{tab:mos} also contains CI (Confidence Interval) of MOS score. As seen in the Table~\ref{tab:mos}, the proposed approach yields significant improvements over the baseline system published by the organizers. In particular, the global MOS score improves by an absolute measure of $0.9$ on the subjective evaluation over the noisy audio. This proposed model involving A-FaSNet based beamforming with the single channel enhancement pipeline also fares well when comparing with the enhanced baseline. The  results are also consistent in other metrics like S-MOS and N-MOS.  

For objective score evaluation, the  perceptual evaluation of speech quality (PESQ) \cite{pesq}, short-time objective intelligibility measure (STOI) \cite{stoi} and scale invariant-signal to noise ratio (SI-SNR) are used as metrics. We would like to note here that many of these objective quality metrics are more suited for noise suppression objective (additive noise) as supposed to reverberation artifacts in far-field speech (convolutive noise). 

The objective quality measurement results are given in Table~\ref{tab:pesq}. Here, we have computed the PESQ score on the development data shared by the organizers. We observe that, with the proposed A-FaSNet model, we obtain some gains in the PESQ score over the noisy signal, and there are further improvements that can be seen when we apply single channel enhancement on that. Also one thing to observe is that, even though our proposed model’s PESQ score is less than the baseline development set which is composed of simulated data, it is performing better than the baseline model on evaluation set based on the subjective scores shared by the organisers. As mentioned previously, eval dataset composed of only real recordings, so we can infer that our proposed system is performing better than the baseline system in dealing with real scenarios.

\subsection{Model Parameters And Real time factor}
FaSNet with Attention Model has $4$ million parameters, while the single channel enhancement residual network has $3.76$\,M parameters totalling to $7.76$\,M parameters.
The the average time required by a Tesla V100 SXM2 with $32$\,GB RAM to process $1$ second audio is $6$\,ms whereas the same for single channel enhancement module is $2.6$\,ms. Thus, the proposed methods are about $100$ times real time.


\section{Conclusions\label{sec:summary}}

The proposed FaSNet with attention gave promising results for beamforming task. This approach combined with the single channel enhancement methods further improved the speech quality. Several objective and subjective evaluation results highlight the benefits from the proposed modeling framework. This research work has been done only for study purpose.

\section{Acknowledgments}\nonumber
We thank the organizers of the ConferencingSpeech 2021 for providing the data, baseline system results and for subjectively evaluating the final system submission and the baseline system. 

\newpage
\bibliographystyle{IEEEtran}

\bibliography{mybib}

\end{document}